\documentclass{article}
\usepackage{arxiv}
\usepackage{amsmath}
\usepackage[utf8]{inputenc} 
\usepackage[T1]{fontenc}    
\usepackage{hyperref}       
\usepackage{url}            
\usepackage{booktabs}       
\usepackage{amsfonts}       
\usepackage{nicefrac}       
\usepackage{microtype}      
\usepackage{lipsum}
\usepackage{graphicx}
\graphicspath{ {./images/} }

\makeatletter
\renewcommand{\@maketitle}{%
  \vbox{%
    \hsize\textwidth
    \linewidth\hsize
    \vskip 0.1in
    \@toptitlebar
    \centering
    {\LARGE\scshape\@title\par}
    \@bottomtitlebar
    \vskip 0.14in
    {\normalfont\large\@author\par}
    \vskip 0.22in
  }%
}
\makeatother

\title{An Integrated Scientific AI Framework for Inorganic Materials Design and Industrial Process Optimization}

\author{
 Yingjie Shi\textsuperscript{*}\quad Yiru Gong\quad Yiqun Su\quad Suya Xiong\quad Jiale Han\textsuperscript{*}\quad Runtian Miao \\[6pt]
 {\normalfont\small\hypersetup{hidelinks}\textsuperscript{*}Correspondence: \href{mailto:yingjies@orachord.com}{yingjies@orachord.com}; \href{mailto:jialeh@ecomatrixai.com}{jialeh@ecomatrixai.com}}
}

\begin{document}
\maketitle
\begin{abstract}
Artificial Intelligence (AI) is redefining the frontiers of scientific domains, ranging from drug discovery\cite{jumper2021highly, blanco2023role, zhang2025artificial, jayatunga2022ai} to meteorological modeling\cite{dewitte2021artificial, conti2024artificial, bi2023accurate}, yet its integration within industrial manufacturing remains nascent and fraught with operational challenges. To bridge this gap, we introduce Aethorix v1.0, an AI agent framework designed to overcome key industrial bottlenecks, demonstrating state-of-the-art performance in materials design innovation and process parameter optimization. Our tool is built upon three pillars: a scientific corpus reasoning engine that streamlines knowledge retrieval and validation, a diffusion-based generative model for zero-shot inverse design, and specialized interatomic potentials that enable faster screening with $ab$ $initio$ fidelity. We demonstrate Aethorix's utility through a real-world cement production case study, confirming its capacity for integration into industrial workflows and its role in revolutionizing the design-make-test-analyze loop while ensuring rigorous manufacturing standards are met.
\end{abstract}


\section{Introduction}

The Fourth Industrial Revolution (Industry 4.0) has ushered in a new era of manufacturing defined by the pervasive integration of cyber-physical systems, artificial intelligence (AI), and digital fabrication.\cite{petrillo2018fourth, philbeck2018fourth} This paradigm enables mass personalization, smart products, and data-driven agility,\cite{schwab2024fourth} yet it has also exposed a critical asymmetry: while digital capabilities advance exponentially, the backbone of (bio)materials innovation remains entrenched in conventional, human-intuition-driven methodologies.\cite{pyzer2022accelerating} This bottleneck is acutely highlighted by the vastness of chemical space; the $>10^{60}$ drug-like compounds, for instance, exceed human cognitive capacity by tens of orders of magnitude.\cite{reymond2012enumeration} While the development of self-driving laboratories represents a significant step forward,\cite{szymanski2023autonomous, dai2024autonomous, pyzer2022accelerating, seifrid2022autonomous} their output often struggles to translate into realistic, manufacturable products, due in part to an overreliance on idealized computational data that fails to capture real-world non-ideality.\cite{peplow2025ai} \\

Interviews with industrial stakeholders across the United States and China reveal that the core challenge lies not only in materials innovation but, more critically, in the manufacturing phase, which are governed by intricate, multi-step, multi-scale thermophysical and chemokinetic phenomena.\cite{zhao2020multiscale, oliveira2024learning} Here, high-fidelity atomistic simulations are computationally prohibitive to deploy at the spatial and temporal scales of continuous production, whereas agnostic data-driven methodologies lack embedded physicochemical realism, rendering them unreliable for predictive control and scale-up due to their failure to generalize beyond calibrated regimes. Bridging this gap requires an integrated framework that synergistically combines operational data with physics-based models.\cite{blakseth2022combining} \\

However, the current deployment of AI in scientific domains remains largely confined to auxiliary tasks such as literature mining, information retrieval, and decontextualized property analysis.\cite{haristiani2019artificial, miret2024llms, kim2024large, cavanagh2024smileyllama, tang2025matterchat,xiao2024proteingpt, pal2023chatgpt, kon2025curie} Achieving higher-order scientific reasoning and innovation is hampered by several core limitations: (1) data fragmentation across material classes impedes cross-domain generalization;\cite{schmidt2019recent} (2) the high-dimensional, context-dependent nature of material behavior, governed by multivariate conditions such as temperature, pressure, and local chemical environment, renders decontextualized predictions scientifically incomplete;\cite{gurnani2024ai} and (3) the multi-scale character of physicochemical phenomena, spanning femtosecond electronic excitations to centennial geological processes, presents a major obstacle for unified modeling.\cite{butler2018machine} \\

An emerging frontier is the transition from passive computational predictive models to active, goal-directed inverse design.\cite{wang2022inverse, zeni2025generative, fuhr2022deep, menon2022generative} This trending methodology would enable the autonomous navigation of chemical space toward desired properties, moving beyond the brute-force, trial-and-error approach of high-throughput screening, which merely accelerates evaluation rather than intelligently guiding exploration.\cite{zeni2025generative} Overcoming these intertwined bottlenecks in both materials innovation and process optimization is critical for vertical industries to meet accelerating market demands and sustain a competitive advantage. \\

Herein, we developed Aethorix v1.0 AI agent for inorganic materials innovation and process optimization, grounded in inverse design principles. The agent is primarily designed to replace the traditional, labor-intensive trial-and-error approaches with direct material candidate generation, enabling the direct identification of novel materials without relying on exhaustive, human-biased screening. Aethorix v1.0 factors natural complexities into its generative design framework, including structural disorder, surface functionalization, surface reconstruction, and temperature-dependent effect. This enables the framework to propose structural modifications tailored to specific target properties under operational physicochemical conditions, thereby further facilitating efficient navigation of complex material spaces to identify high-performance inorganic compounds. Another cornerstone of Aethorix v1.0 is its accelerated property prediction engine, which delivers first-principles accuracy at speeds compatible with industrial production timelines. To demonstrate the superiority of our agent over traditional industrial workflows, we will present a use case focused on enhancing cement production quality.

\section{Aethorix v1.0 Workflow}

\begin{figure} [htbp]
    \centering
    \includegraphics[width=1\linewidth]{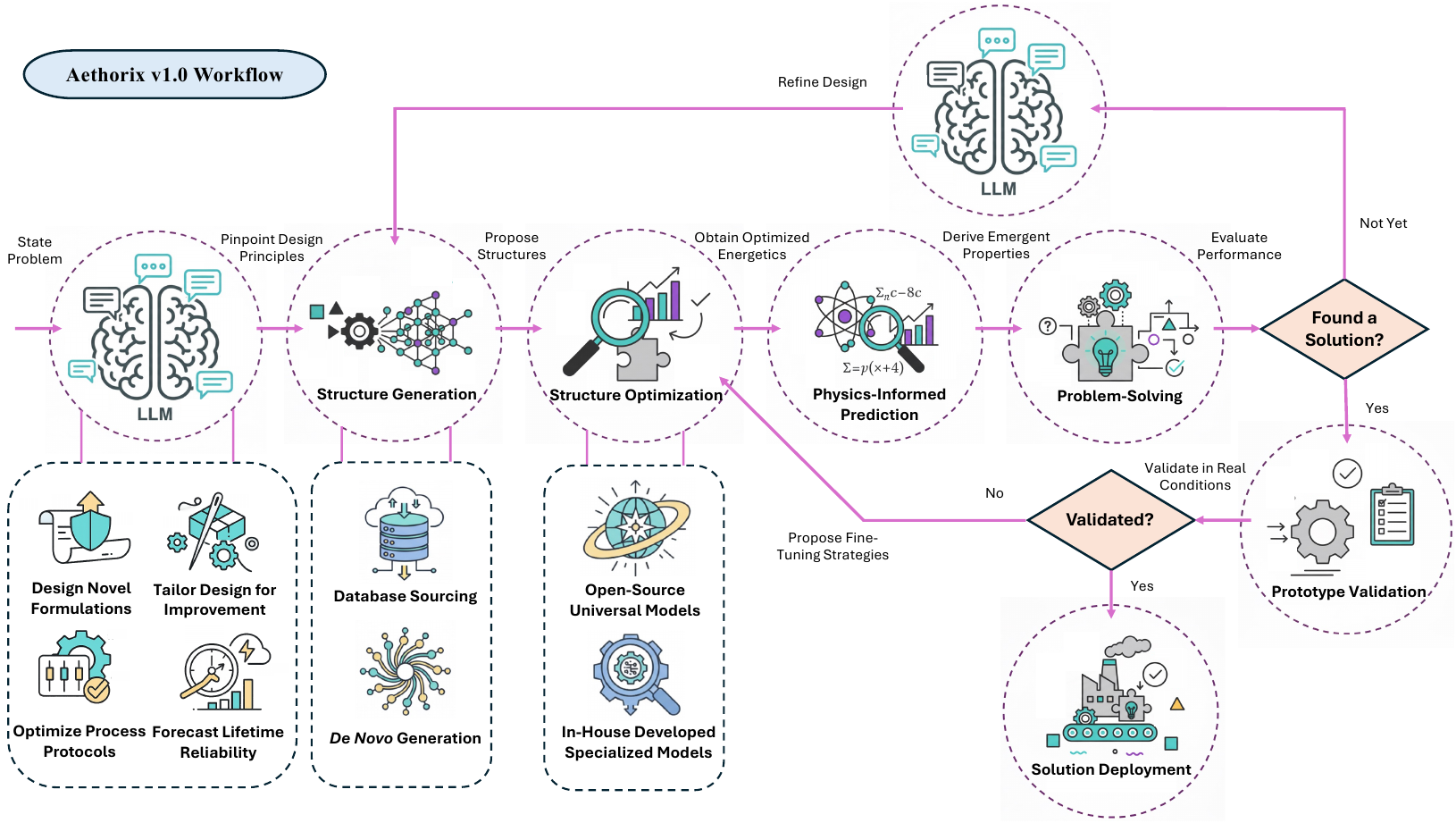}
    \caption{\textbf{Schematic Illustration of the Aethorix v1.0 Workflow and Its Industrial Applications.} Aethorix v1.0 integrates LLM-guided problem definition, generative structure design, structure optimization, and physics-informed prediction into an iterative loop for solving industrial challenges. Candidate solutions are refined until validated under real-world conditions, enabling prototype testing and industrial deployment. The scientific AI agent underpins applications including but not limited to novel material formulations design, tailored material design improvement, process protocol optimization, and lifetime reliability forecasting. } 
    \label{main}
\end{figure}

Our scientific AI agent executes a closed-loop, data-driven, and physics-embedded inverse design paradigm for the automatic, zero-shot design of material formulations and optimization of industrial protocols. The main workflow is shown in Figure \ref{main}. The Aethorix v1.0 agent initiates its operational cycle by performing abstract problem decomposition and ontological reasoning. It leverages its Large Language Model (LLM) module to conduct exhaustive multimodal analysis, identifying established knowledge, precedent, and critical research gaps within the target scientific domain. This step formalizes a defined industrial challenge into a structured set of design principles and constraints ($e.g.$ chemical space, operational environmental conditions). Subsequently, the agent invokes its Structure Generation module to propose an atomistic manifold of novel, unique, and stable candidate structures that fulfill the specified compositional and structural requirements. The generated structures are geometrically optimized by the Structure Optimization module to obtain ground-state thermodynamic properties, such as phase stability and synthetic viability. Next, physical-informed prediction can be applied to calculate emergent macroscale properties, such as electronic, magnetic, thermal, and mechanical characteristics of materials. It is at this stage that the agent performs Problem-Solving, integrating the calculated results with scientific reasoning to derive meaningful insights. \\

The core of the agent's capability is its iterative refinement loop, which actively incorporates feedback from both computational screening results and real-world prototype validation. When a candidate solution fails, the agent performs causal analysis to pinpoint the source of discrepancy and proposes targeted modifications. If the issue stems from a lack of viable candidates, the agent employs its LLM module for causal reasoning to rationally expand or reconfigure the chemical design space. Conversely, if the failure occurs during experimental prototyping, the agent strategically fine-tunes its predictive models to improve the accuracy of future property forecasts. This adaptive process iterates recursively until a candidate solution satisfies all validation thresholds. The converged solution is then elevated to administrative approval and subsequent industrial deployment.

\section{Applications of Aethorix v1.0}

We demonstrate the multifarious capabilities of our AI agent in addressing the complex challenges of the inorganic materials industry through its integrated workflow (See Figure \ref{main}). The agent can be tasked to zero-to-one design new materials formulations and configurations, proposing verified candidates whose compositional spaces and crystallographic symmetries extend beyond any known experimental or computational precedents. This could markedly accelerating the traditionally slow discovery-to-commercialization pipeline. For more common uses, the agent can tailor an existing design to explore elevated functionalities. This typically involves introduction of dopant elements or defects, surface functionalization, and microstructural modification. The overall workflow substantially outperforms conventional trial-and-error approaches by leveraging generative design to constrain the vast chemical search space, followed by accelerated physics-based sampling. \\

Beyond materials design innovation, the agent can also be tasked to optimize industrial process protocols by reverse-engineering process parameters ($e.g.,$ synthesis parameters, thermal treatments, processing conditions) to maximize production efficiency and product quality. The workflow can be applied to partially unpack the "black-box" nature of complex reaction mechanisms by, for example, deducing the possible crystalline polymorphic phases under specific processing conditions, thereby generating synthetic data to augment downstream model training and experimental design. There, parametric optimization can be achieved through Bayesian optimization, which navigates the high-dimensional parameter space to identify optimal processing conditions with minimal experimental iterations. Another important use of our agent is to forecast lifetime reliability through generative modeling of degradation pathways. This is achieved by using the generative framework to explore microstructural evolution sequences that give rise to degradation phenomena, fatigue behavior, and failure modes under operational conditions. By simulating defect propagation, crack initiation, and phase transformation kinetics across multiple timescales, critical failure precursors can be identified. This capability enables proactive maintenance scheduling, optimized warranty period determination, and evidence-based regulatory compliance reporting.

\section{Multi-Scale Dataset}

\subsection{First-Principles Atomistic Data}

The foundational atomistic dataset is constructed from high-fidelity, first-principles calculations based on the density functional theory (DFT). These are the computational ground truth for mapping structure-energy relationships, which are fundamental to the generative and predictive modules within the agent workflow. Our work leverages and extends large-scale, publicly available benchmarks. This includes the Open Molecule 2025 (OMol 25) \cite{levine2025open} (100M+ single-point calculations), the Open Materials 2024 (OMat24) \cite{barroso2024open} (110M+ single-point calculations), and the Open Molecular Crystals 2025 (OMC25) \cite{gharakhanyan2025open} (27M+ single-point calculations) datasets as our primary bulk structure repositories. For surface-specific phenomena, we utilize the Open Catalyst 2020 (OC20) \cite{chanussot2021open} ($\sim$265M single-point calculations) and the Open Catalyst 2022 (OC22) \cite{tran2023open} ($\sim$9.8M single-point calculations) datasets. All datasets consist of an ensemble of unrelaxed and relaxed structures, spanning a wide range of atom types, along with their corresponding DFT-calculated total energies, atomic forces, and stress tensors across diverse adsorbate species. Additional datasets are supplied via our Structure Generator and in-house DFT calculations for specific systems of industrial interest, fulfilling the undersampling gaps for targeted compositions and configurations within public repositories. To ensure data consistency, the exchange-correlation functionals and pseudopotentials used for these supplemental calculations are meticulously aligned with those of the referenced public datasets. All in-house calculations are performed with the Quantum Espresso software package\cite{giannozzi2009quantum, giannozzi2017advanced} and the Vienna $ab$ $initio$ simulation package (VASP)\cite{hafner2008ab}. 

\subsection{Industrial Operational Data}

The industrial-scale data referenced in this work is sourced directly from active manufacturing facilities, capturing the end-to-end process chain for complex, multi-phasic inorganic material production. These datasets are typically composed of high-dimensional, time-series records that log operational parameters, input material specifications, and resultant product quality metrics. Data acquisition occurs through a combination of supervisory control and data acquisition (SCADA) systems for continuous process variables ($e.g.$, feed rates, temperatures) and rigorous laboratory quality control for discrete material characterization. The data structure provides the empirical ground truth for the real-world problem space. Within our agent's workflow, these datasets serve a dual purpose: they can provide the validation benchmark for verifying the real-world performance of computationally designed candidates, while they can also serve as the initial input parameters and target specifications that define the industrial problem to be solved.

\section{Scientific Literature Review}

The scientific literature review is implemented via a Retrieval-Augmented Generation (RAG) pipeline architected around a Large Language Model (LLM). This module ingests pre-processed user queries to conduct multi-stage, vector-based retrieval from a curated scientific corpus (enabled by SciSpace\cite{jain2024scispace}). The retrieval subsystem utilizes open-source, domain-specific text representation models to execute parallel dense searches against a low-latency vector database. This is powered by SPECTER\cite{cohan2020specter} for document-level embeddings and ColBERT\cite{santhanam2021colbertv2} for late-interaction, contextualized token embeddings, which collectively yielding an initial candidate set of several hundred documents. The candidate set is subsequently refined by a neural re-ranker that optimizes for relevance using a composite of features, narrowing the results to the top-k most pertinent passages. The LLM then synthesizes these retrieved, citation-grounded contexts to generate a coherent literature review, thereby constraining generative hallucinations and ensuring the output's verifiability against the source corpus.

\section{Structure Generation}

All atomistic structural modeling is managed by the Atomic Simulation Environment (ASE) \cite{larsen2017atomic}. The source of initial structures is contingent upon the design objective: for targeted exploration of known crystal systems, initial configurations are sourced from the Materials Project\cite{jain2013commentary} database with rational compositional and configurational modifications, whereas for $de$ $novo$ design, the goal is accomplished using a symmetry-aware, diffusion-based generative model. The implementation is adapted from MatterGen\cite{zeni2025generative}, which is architected for zero-shot generation of novel and thermodynamically stable inorganic materials. In this scheme, periodic crystal materials are represented by $M=(A, X, L)$, where $A=(a^1,a^2,...,a^n)^T \in \mathbb{A}^n$ represents the atom species inside the unit cell, $X=(x^1,x^2,...,x^n) \in [0,1)^{3 \times n}$ denotes the fractional coordinates of atoms in the unit cell, and $L=(l^1,l^2,l^3) \in \mathbb{R}^{3 \times 3}$ is the lattice with three vector components. During the forward diffusion (noising) process, $A$, $X$, and $L$ are independently corrupted toward a physically meaningful prior distribution as follows: 

\vspace{-10pt}
\begin{equation}
    q(A_{t+1}, X_{t+1}, L_{t+1} | A_t, X_t, L_t) = q(A_{t+1} | A_t) \, q(X_{t+1} | X_t) \, q(L_{t+1} | L_t)
\end{equation}

\noindent where $q$ denotes distributions and $t$ is the diffusion timestep. The model generates crystal structures by iteratively denoise a random initial structure through learned reverse diffusion. Atom types are corrected through discrete categorical transitions that minimizes the Kullback-Leibler divergence and cross-entropy loss, fractional coordinates are updated using periodic score predictions to approximate the log-probability density, and the lattice matrix is adjusted via equivariant operations that preserve volume and symmetry. The total training loss is the weighted sum of the score matching loss across all parameters, expressed as:

\vspace{-10pt}
\begin{equation}
    \mathcal{L} = \lambda_{\text{coord}}\mathcal{L}_{\text{coord}} + \lambda_{\text{cell}}\mathcal{L}_{\text{cell}}+ \lambda_{\text{types}}\mathcal{L}_{\text{types}}
\end{equation}

To enhance the efficiency of the downstream high-throughput screening process, a property-guided generation capability is integrated. This is achieved by fine-tuning the unconditional score network by augmenting its node representations with property-specific features through trainable adapter modules.\cite{dunn2020benchmarking} The property-augmented node hidden representation is given by:

\vspace{-10pt}
\begin{equation}
    H_j'^{(L)} = H_j^{(L)} + f^{(L)}_{\text{mixin}}\Big(f^{(L)}_{\text{adapter}}(\mathbf{g})\Big)
\end{equation}

\noindent where f$^{(L)}_{\text{mixin}}$ and f$^{(L)}_{\text{adapter}}$ are the $L$-th mix-in and adapter layer, respectively. The base MatterGen framework supports conditioning on properties including chemical system, space group, magnetic moment, band gap, bulk modulus, and energy above hull. We are actively extending this conditioning space to encompass a broader range of user-specified performance metrics, with a particular focus on multi-property conditioning to generate structures that concurrently satisfy multiple targeted requirements. We have also developed fine-tuned models in the crystal structure prediction (CSP) mode, where the model denoises only the atomic coordinates and lattice not atom types, preserving the stoichiometry of the input composition. This is especially valuable for generation of fixed stoichiometric, polymorphic systems, as it enables efficient and exhaustive sampling of the configurational not compositional landscape. 

\section{Structure Optimization}

DFT is a standard tool for the ground-state energetics calculations of crystal structures, which can be followed by geometric optimization to determine the equilibrium atomic configuration. This is typically achieved using the limited memory Broyden-Fletcher-Goldfarb-Shanno (L-BFGS)\cite{liu1989limited} or the fast inertial relaxation engine (FIRE)\cite{bitzek2006structural} algorithm. However, its computational cost makes it difficult to scale for high-throughput industrial applications, necessitating the development of surrogate models that reproduce $ab$ $initio$ accuracy at orders-of-magnitude lower computational expense. Machine-learned interatomic potentials (MLIPs) utilize parametric functionals to represent interatomic interactions. Trained on atomic structure data, MLIPs are developed through a graph neural network (GNN)-based architecture, which directly maps atomic configurations to their corresponding potential energy surface (PES), without the need for handcrafted feature descriptors. Many state-of-the-art (SOTA) GNNs, such as MACE\cite{batatia2022mace}, NequIP\cite{batzner20223}, eSEN\cite{fu2025learning}, and PaiNN\cite{schutt2021equivariant}, have been developed to leverage equivariant message passing to achieve high data efficiency. Our agent integrates the open-source Universal Models of Atoms–Medium (UMA-M)\cite{wood2025family}, which comprises five specialized models, each uniquely designed for distinct categories of material systems. While the models have exhibited remarkable generalizability across a large chemical space, prediction accuracy for novel compositions may still be constrained by the inherent limitations of the training data, which lack a proper description of charge, spin, and strong $d$-orbital correlations. For specialized systems where these effects are dominant, our agent develops specialized MLIPs, which are typically trained on only a few atom types, but they can be rapidly fine-tuned and uncertainty quantified via ensemble variance methods. Training of MLIPs is implemented in a multi-target learning scheme, where the node features are optimized to simultaneously predict multiple properties. The loss function is given by:

\vspace{-10pt}

\begin{equation}
    \mathcal{L} = \frac{1}{N} \sum_{i=1}^{N} \left[
    \alpha_E \left\| E_i - \hat{E}_i \right\|^2 +
    \alpha_F \left\| \mathbf{F}_i - \hat{\mathbf{F}}_i \right\|^2 +
    \alpha_S \left\| \boldsymbol{\sigma}_i - \hat{\boldsymbol{\sigma}}_i \right\|^2
    \right]
\end{equation}

\noindent where $\alpha_{E}$, $\alpha_{F}$, and $\alpha_{S}$ are the weight coefficients for total energy, atomic force, and virial stress, respectively, $N$ denotes the number of training samples, $X_i$ and $\hat{X}_i$ ($X=E, \mathbf{F}, \sigma$) are the true and predicted targets for structure $i$, respectively. Optimized hyperparameters are often determined using Bayesian optimization methods.\cite{martinez2018practical} MLIP ensembles are constructed by training multiple models on identical datasets but with different parameter initializations. The variance in predictions across the ensemble provides a rigorous, $a$ $priori$ metric for assessing model confidence and identifying extrapolation beyond the trained domain\cite{musil2019fast}. Active learning is implemented to ensure transferability of specialized models across the desired chemical space.\cite{podryabinkin2017active, podryabinkin2019accelerating, zhang2019active} The training dataset is iteratively enriched by incorporating new structures that are generated from the conditioned atom types and have the highest epistemic uncertainty.

\section{Physics-Informed Prediction}

\begin{figure} [htbp]
    \centering
    \includegraphics[width=1\linewidth]{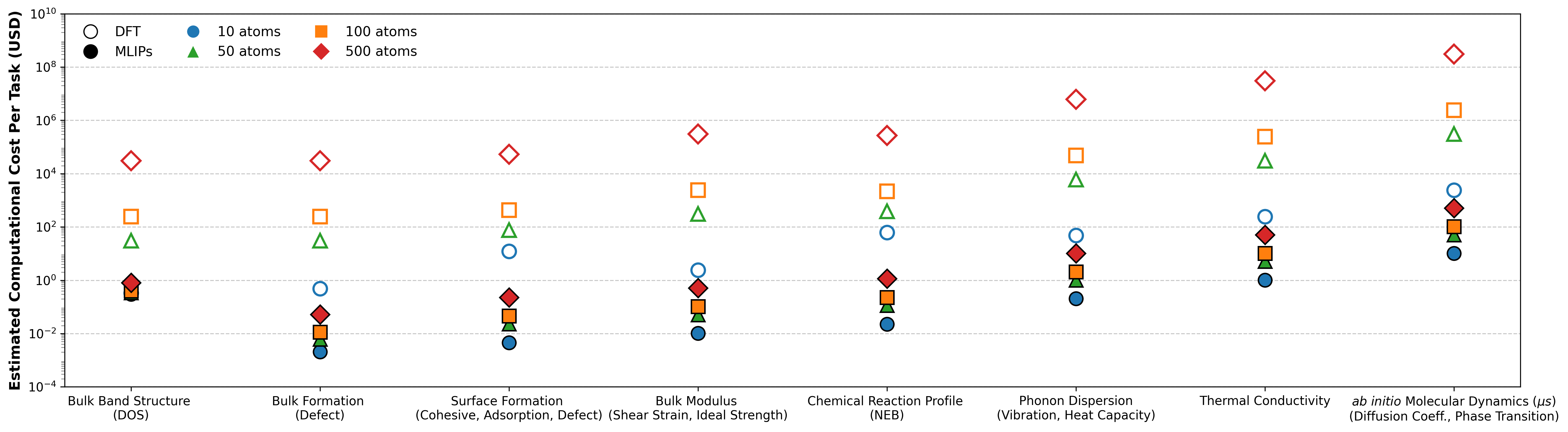}
    \caption{\textbf{Computational Cost Comparison for Common Materials Emergent Properties.} Hollow and solid markers denote tasks completed using DFT and MLIP, respectively. MLIP shows 3–5 orders of magnitude greater computational economy compared to DFT for deriving emergent properties. Costs of DFT calculations and MLIP inferences were benchmarked on an Intel Xeon Platinum 8175M CPU (32 cores) and on a single NVIDIA Tesla T4 GPU, respectively. All hardware instances were provisioned through Amazon Web Services. } 
    \label{cost}
\end{figure}

Many emergent macroscale properties of materials cannot be simply derived from a single DFT geometric optimization calculation, but instead require multiple DFT calculations integrated with physics-informed analysis. Hence, the computational cost can grow exponentially with increasing system size and project scope. In contrast, MLIPs can unlock the unprecedented efficiency in predictive screening while preserving near-$ab$ $initio$ fidelity. To enable higher-level property predictions, our current version provides direct interfaces to Pymatgen\cite{ong2013python}, Phonopy, Phono3py\cite{togo2023first, togo2023implementation}, and DeePMD-kit.\cite{zeng2023deepmd} Figure \ref{cost} contrasts the computational cost scaling between DFT and MLIPs across a range of system sizes for various simulation tasks. These performance benchmarks are averaged from the estimated calculation convergence time observed across a series of Mg-Si-O-H chemical systems. The computational cost of obtaining emergent properties via MLIP inference is 10$^3$ $\sim$ 10$^5$ times lower than equivalent DFT calculations. It should be noted that the exact magnitude of this advantage can vary with the chemical system, the resolution of the computational methods employed ($e.g.,$ the number of intermediate images in nudged elastic band calculations), and the simulation timescale ($e.g.,$ for molecular dynamics trajectories), yet it consistently reflects a significant reduction in computational expense for MLIP. Crucially, this cost reduction renders many prohibitively expensive calculations ($e.g.,$ phonon vibrations) feasible for high-throughput screening.

\section{Use Case}

\subsection{Problem Framing}

\begin{figure} [htbp]
    \centering
    \includegraphics[width=1\linewidth]{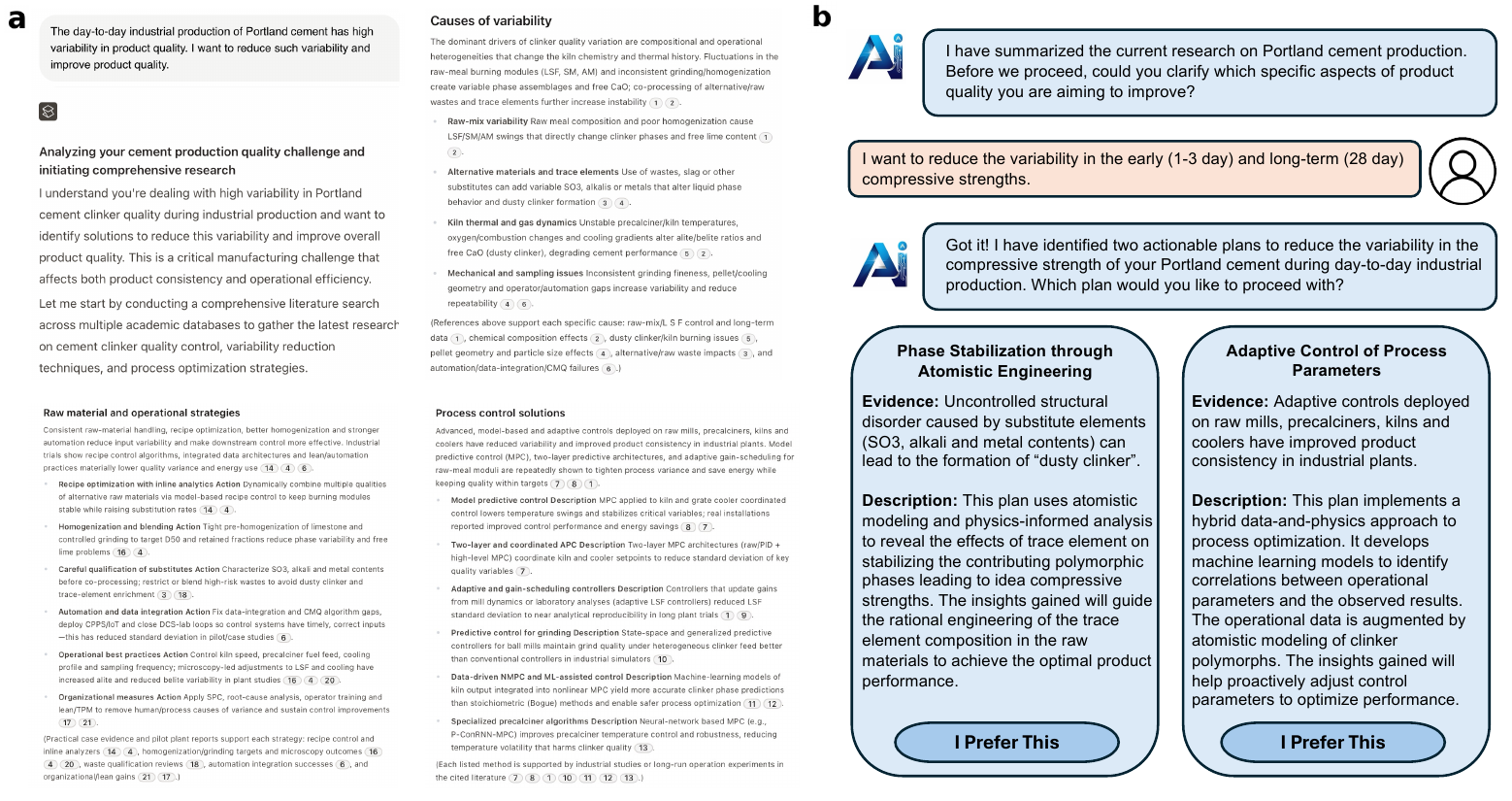}
    \caption{\textbf{Translating Problem Statements into Design Principles.} \textbf{a,} Knowledge synthesis of scientific literature by the integrated SciSpace module, identifying problem root causes through evidence-based reasoning and proposing solution strategies based on the agent's capability. \textbf{b,} Solution formulation facilitated by the Aethorix v1.0 agent, where user interaction enables the development and iterative refinement of an actionable implementation plan. The agent's responses (blue) and user's feedback (orange) are color-coded.} 
    \label{llm}
\end{figure}

Herein, we present a use case of Aethorix v1.0 in resolving a persistent quality control challenge in Portland cement manufacturing. As detailed in Figure \ref{llm}, the objective of the example use case is to systematically reduce the day-to-day production variability in the compressive strength across both early-age (1-3 day) and long-term (28 day) curing regimes. After consolidating the predominant expert perspectives from a corpus of 18 scholarly works published within the last five years,\cite{fayaz2025industrial, venkatesh2020predictive, pollabauer2024modellbasierte, zanoli2020optimization, tsamatsoulis2024robust, pan2024operation, lyu2024quality, hao4590797control, ali2022machine, teplicka2023evaluation, sutawijaya2021optimizing, farazmand2024effect, saied2020process, costa2020portland, kolesnikov2021review, li2022substance, huilcapi2020analisis, masmali2021implementation} the integrated SciSpace module summarized leading causal factors and proposed established mitigation strategies for raw materials engineering and process control optimization (Figure \ref{llm}a). Following user interaction, our agent formulated actionable plans for solution development (Figure \ref{llm}b), prompting the user to select an implementation path based on their prioritized needs. It is noteworthy that the framework supported further interactive dialogue to elaborate on the underlying design principles of each proposed solution.

\subsection{Phase Stabilization through Atomistic Engineering}

\begin{figure} [htbp]
    \centering
    \includegraphics[width=1\linewidth]{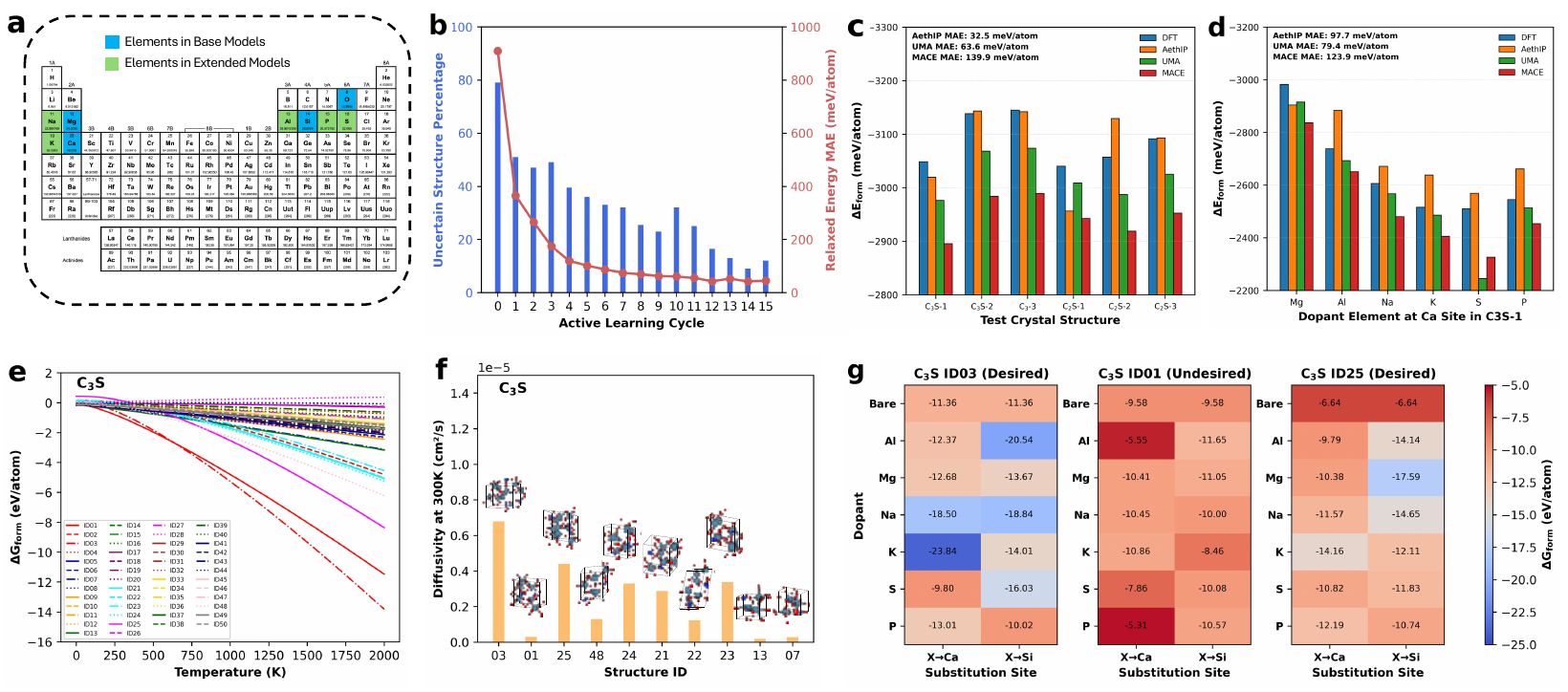}
    \caption{\textbf{Trace Element Stabilization of C$_3$S Clinker Phases.} \textbf{a,} Elemental scope of the customized AethIP potential. The base model was trained on the Ca-Mg-Si-O system of the host structure, while extended models were fine-tuned to include specific dopant elements. \textbf{b,} Active learning history of the base AethIP model. Model transferability was assessed on-the-fly by quantifying the number of uncertain structures within newly generated ensembles. A structure was flagged as uncertain if the force standard deviation $>$ 1 eV/\AA{} and/or energy standard deviation $>$ 40 meV/atom. \textbf{c,} Transferability test of base AethIP on three unseen C$_3$S and C$_2$S polymorphs, respectively. AethIP significantly outperforms both UMA-M-OC20 and MACE-MPA-0 for structural optimization and formation energy prediction. \textbf{d,} Transferability test of extended AethIP on six doped C$_3$S single phase. AethIP exceeds the accuracy of MACE-MPA-0 but not that of UMA-M-OC20. \textbf{e,} Gibbs free energies of formation as a function of temperature (0–2000 K) for the 50 most stable C$_3$S polymorphic phases identified at 0 K. \textbf{f,} Ionic self-diffusivities at 300 K for the ten most stable C$_3$S polymorphs synthesized at 1730 K. Lattice structures are visualized above the corresponding bars, ordered by descending thermodynamic stability at 1730 K. \textbf{g,} Heat map of the formation Gibbs free energy change induced by 10 mol\% dopant substitution at each crystallographic site in the most stable C$_3$S polymorphs synthesized at 1730 K. The color scale represents the formation energy difference relative to the bare (undoped) structure at 0 K. }
    \label{phase}
\end{figure}

All computational simulation results used to derive the phase stabilization solution are presented in Figure \ref{phase}, including the in-house developed specialized MLIPs (denoted AethIP) employed for optimizing all clinker polymorphic structures (a-d). The base AethIP models were developed using a PaiNN architecture and trained over 15 AL cycles (Figure \ref{phase}a,b). Each cycle iteratively incorporated new Ca-Mg-Si-O configurations selected for maximizing model uncertainty, based on force and energy prediction standard deviation. Subsequently, models were extended to cover additional elements, including Na, Al, P, S, and K using the same methods. The final AethIP was tested on a few Ca$_3$SiO$_5$ (C$_3$S) and Ca$_2$SiO$_4$ (C$_2$S) polymorphs across a range of crystalline symmetries (Figure \ref{phase}c), as well as against a single C$_3$S polymorph containing different dopants (Figure \ref{phase}d). AethIP outperformed both UMA and MACE pretrained universal MLIPs for pristine structure optimization and formation energy prediction, achieving an MAE of 32.5 meV/atom for final energy. For doped structures, AethIP also achieved lower MAE than MACE. It came to our notice that our training motif worked very well for base models but the extended models show diminished performance when tested on new elements. The performance gap highlights an area for future research and methodological refinement. \\

The elevated calcination temperature of 1730 K necessitated explicit consideration of entropic contributions to phase stability. C$_3$S and C$_2$S are the dominant compositions in cement clinkers. The computed Gibbs free energies of formation for polymorphic C$_3$S and C$_2$S phases at finite temperatures are presented in Figure \ref{phase}e–g and Figure \ref{Sphase}, respectively. The thermodynamic landscape reveals a broad distribution of triclinic and orthorhombic symmetries across both systems. The distribution shifted significantly between room temperature and 1730 K. While experimental identification of exact C$_3$S and C$_2$S crystallographic symmetries remains challenging, our predicted most stable phases exhibit good consistency with the reported industrial clinker polymorphs.\cite{ludwig2015research} At 1730 K, the equilibrium lattice parameters of the most thermodynamically stable C$_3$S and C$_2$S polymorphs are shown as follows:

\begin{table}[h!]
\centering
\caption{Lattice parameters for the most stable pristine C\(_3\)S and C\(_2\)S polymorphs identified at 1730 K.}
\label{tab:lattice_params}
\begin{tabular}{lcccccc}
\hline
\textbf{Phase} & \textbf{a (\AA)} & \textbf{b (\AA)} & \textbf{c (\AA)} & \boldmath$\alpha$ (\textbf{°}) & \boldmath$\beta$ (\textbf{°}) & \boldmath$\gamma$ (\textbf{°}) \\
\hline
C\(_3\)S & 12.68 & 17.46 & 13.78 & 87.53 & 87.77 & 91.81 \\
C\(_2\)S & 10.59 & 9.93 & 14.89 & 111.38 & 98.02 & 92.46 \\
\hline
\end{tabular}
\end{table}
The early-age properties and long-term performance of concrete are directly governed by the hydration behavior of its clinkers.\cite{claverie2020ab, geng2025investigation} However, the hydration of clinkers involves complex, coupled reaction mechanisms ($e.g.,$ ionic dissolution, degree of hydroxylation, evolution of surface structure, and chemical bond formation),\cite{geng2025investigation, chen2025molecular} making the modeling of the full process at atomistic levels prohibitively expensive. Our agent captured the bulk ionic mobility to approximate hydration reactivity of clinkers. Figure \ref{phase}f shows the ionic self-diffusivity distribution for the thermodynamically most stable C$_3$S polymorphs, quenched from 1730 K. The self-diffusivity values were derived from short-time molecular dynamics simulations (1 ps) at 300 K via the Einstein relation:

\vspace{-10pt}
\begin{equation}
    D = \lim_{t \to \infty} \frac{1}{2d} \frac{d}{dt} \left\langle \lvert \mathbf{r}_i(t) - \mathbf{r}_i(0) \rvert^2 \right\rangle
\end{equation}

\noindent where $N$ is the number of atoms, $d$ is the dimensionality of diffusion, $\mathbf{r}_i$ is the position of atom $i$ at $t$, and the
angular brackets denote an ensemble or time average. The self-diffusivity was found to vary by 1–2 orders of magnitude across all thermodynamically most stable polymorphs. Those with low ionic mobility ($e.g.,$ ID 01) were hence identified undesirable and their production was targeted for suppression during synthesis. Figure \ref{phase}g and S1c show the phase stabilization effect due to presence of low concentration ($\sim$10 mol\%) of dopant ions in clinker crystal lattices at 1730 K. The results revealed varying dopant site occupancy preferences across polymorphs, which deviate from empirical trends.\cite{zhang2011development, young1996highly, schwiete1968existenzbereiche} For instance, Na preferentially substitutes the Ca site in the C$_2$S ID10 lattice, whereas it favors the Si site in the C$_3$S ID25 lattice. Most of these substitutions resulted in a reduced free energy at 1730 K, Notable exceptions, such as S or P substituting for Ca, are likely due to the significant charge mismatch introduced by the dopant, which destabilizes the lattice. These are supported by experimental observations.\cite{ludwig2015research} Crucially, K doping provided the strongest stabilization of all dopants for the desired polymorphs, while having a negligible effect on the undesired ones. It was further verified that incorporation of K does not impede ion mobility within the desired phases. Therefore, it was proposed that industrial clinker production strategically maintain K impurity levels at 5–10 mol\% to favor the formation of hydration-reactive polymorphs, a direct route to improving the performance of concrete from early ages to long-term service.

\subsection{Adaptive Control of Process Parameters}

\begin{figure}[htbp]
    \centering
    \includegraphics[width=1\linewidth]{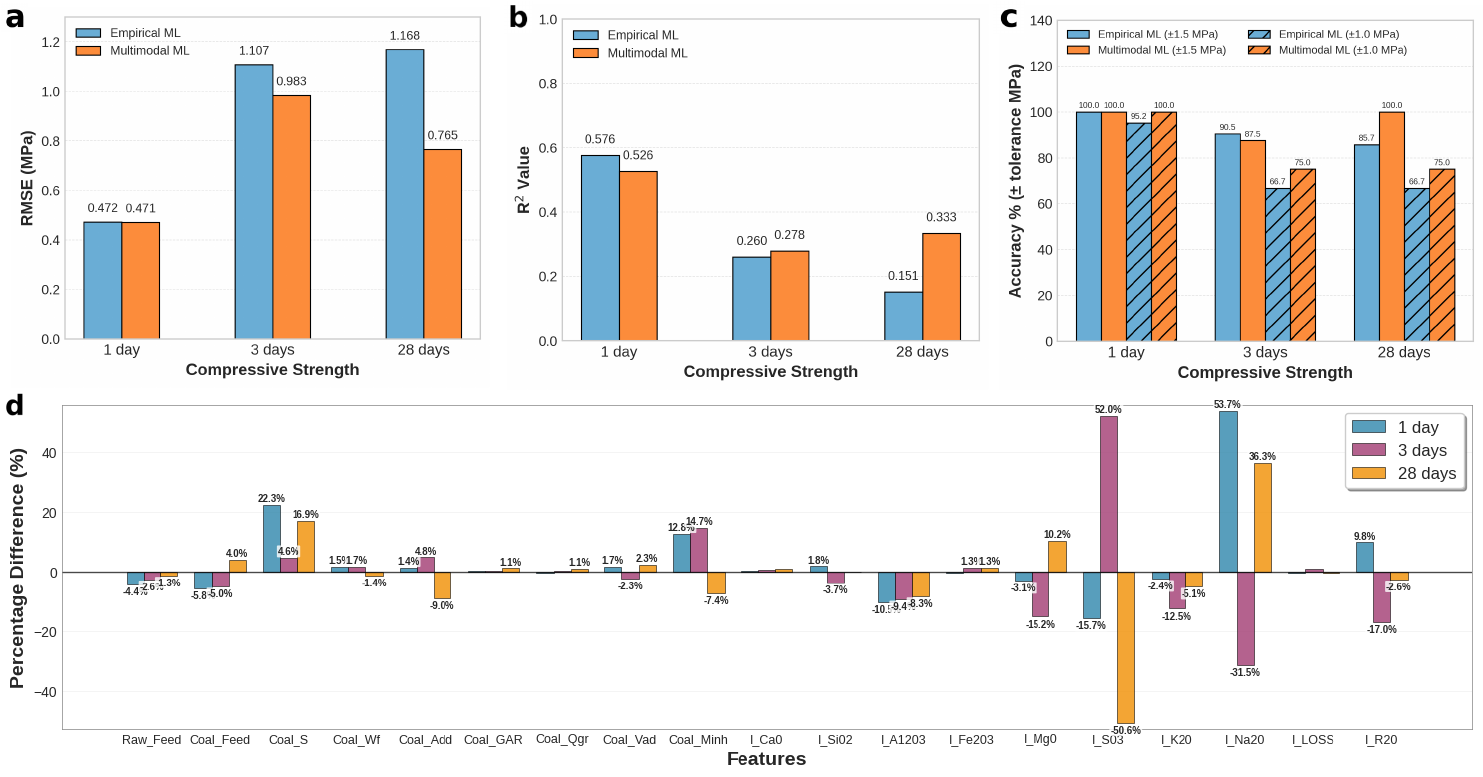}
    \caption{\textbf{Enhanced Cement Compressive Strength Prediction via Multimodal ML.} Performance comparision between empirical, purely data-driven and multimodal ML methods, evaluated through \textbf{a,} root mean squared error (RMSE), \textbf{b,} coefficient of determination ($R^2$), and \textbf{c,} prediction accuracy within thresholds of $\pm 1.5$ MPa and $\pm 1.0$ MPa. \textbf{d,} Optimized parameter deviations from operational means for compressive strength maximization across curing periods, expressed as percentage differences. Deviations less than 1\% are not shown on the plot. }
    \label{control}
\end{figure}

We obtained a comprehensive dataset comprising 82 industrial operational records for year 2025, representing batch production outcomes from cement manufacturing processes. This shows an example how our users can use Aethorix v1.0 to develop predictive models from their proprietary, limited-volume industrial data. ML models were developed to model the complex nonlinear relationships between cement production parameters and resultant compressive strength. Figure \ref{control}a-c shows a comparative analysis between empirical, purely data-driven ML models and our proprietary multimodal approaches. All model development employed a structured data partitioning scheme with a 7:2:1 ratio for training, validation, and test sets, respectively. Each training and validation was repeated for five times to obtain the best result. All ML training protocols detailed in this use case involved a preprocessing step of feature selection prior to implementation of Extreme Gradient Boosting (XGBoost).

The high-quality data scarcity typical of industrial contexts can yield an ill-posed inverse problem for model training. In our example, the empirical models trained on the small dataset exhibited reasonable accuracy for forecasting 1-day compressive strength, but showed significant root mean squared errors (RMSEs) for 3-day and 28-day results. In contrast, the multimodal models reduced the RMSEs by 20-30\% for 3-day and 28-day models, elevating all conservative tolerance threshold ($\pm$ 1.5 MPa) accuracy to above 85\% and stringent precision threshold ($\pm$ 1.0 MPa) accuracy to above 75\%. The multimodel methods incorporated graph representations of predicted dominant clinker and post-hydrated gel phases, as well as scalar value of the corresponding formation energy at 0 K for each phase. The elemental reference energy was shifted in accordance to the stoichiometric ratios of input oxides, as given by:

\vspace{-10pt}
\begin{equation}
    \Delta \mu_{el} = k_B T \ln \left( \frac{x_{el}}{x_{Ca}} \right)
\end{equation}

\noindent where $x_{el}$ and $x_{Ca}$ denote the molar compositions of element $x$ and Ca, respectively. The formalism rests on the assumption of ideal solution behavior. Upon completing model training, Bayesian optimization was employed to find process parameters that maximize cement compressive strength across all curing periods (1-day, 3-day, and 28-day) by iteratively refining significant parameters within physically realistic bounds. The optimized parameters identified for maximizing compressive strength are shown in Figure \ref{control}d, which quantifies the percentage deviation of these parameter values relative to the baseline operational averages employed in the past production. Strategic production decisions for future production can be derived from these optimization results. As examples, systematic reduction of Al content demonstrates potential for quality enhancement across all curing periods, while S content requires careful balancing. Elevated concentrations benefit 3-day strength development but inversely impact 28-day performance. It is worth noting that Figure \ref{control}d shows that the maximum strength would be achieved with lowered K content across all curing periods, contrary to atomistic predictions of phase stabilization through K-doping. The discrepancy may originate from limitations in the MLIP's accuracy for dopant modeling (see Figure\ref{phase}d) or from kinetic barriers, such as inhibited K transport during clinker formation, which prevent the realization of thermodynamically stable phases. While this necessitates iterative refinement of our proprietary predictive models, a detailed discussion of further development falls beyond the scope of this paper.

\section{Conclusion}

The development and validation of Aethorix v1.0 demonstrate a significant stride toward closing the gap between AI-driven scientific innovation and industrial manufacturing. The industrial case study on cement production provides concrete validation of its functionality. Aethorix v1.0 agent successfully navigated the problem from root-cause analysis to actionable solutions, identifying atomistic solution for phase stabilization and providing optimized process parameters through multimodal ML and Bayesian optimization. For future development, we envision extending Aethorix's generative and reasoning capabilities to a broader class of complex, multi-phase systems and dynamic process control challenges, ultimately establishing a new paradigm for the scalable deployment of AI across the entire materials and manufacturing value chain.

\bibliographystyle{unsrt}  
\bibliography{references}

\newpage
\section*{Appendix}
\renewcommand{\thefigure}{S\arabic{figure}}
\setcounter{figure}{0}
\begin{figure} [htbp]
    \centering
    \includegraphics[width=1\linewidth]{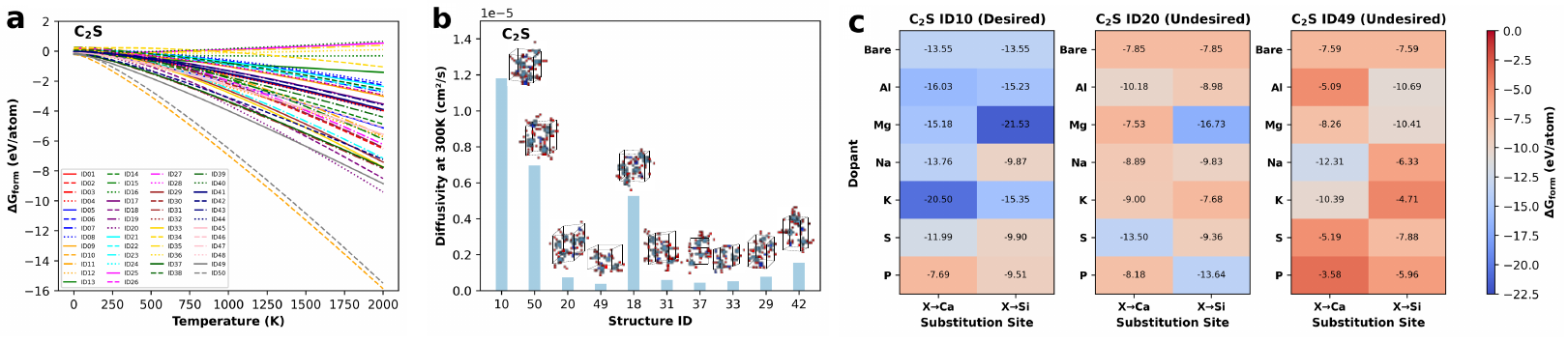}
    \caption{\textbf{Trace Element Stabilization of Clinker C$_2$S Phases.} \textbf{a,} Gibbs free energies of formation as a function of temperature (0–2000 K) for the 50 most stable C$_2$S polymorphic phases identified at 0 K. \textbf{b,} Ionic self-diffusivities at 300 K for the ten most stable C$_2$S polymorphs synthesized at 1730 K. Lattice structures are visualized above the corresponding bars, ordered by descending thermodynamic stability at 1730 K. \textbf{c,} Heat map of the formation Gibbs free energy change induced by 10 mol\% dopant substitution at each crystallographic site in the most stable C$_2$S polymorphs synthesized at 1730 K. The color scale represents the formation energy difference relative to the bare (undoped) structure at 0 K.} 
    \label{Sphase}
\end{figure}

\end{document}